\documentclass[aps,prl,twocolumn,showpacs,superscriptaddress]{revtex4}
\usepackage{graphicx}
\usepackage{verbatim}
\usepackage{sidecap}

\begin{document}

\title{Observation of intertwined Fermi surface topology, orbital parity symmetries and electronic interactions in iron arsenide superconductors}



\author{L.A. Wray}
\affiliation{Department of Physics, Joseph Henry Laboratories of
Physics, Princeton University, Princeton, NJ 08544, USA}
\author{D. Hsieh}
\affiliation{Department of Physics, Joseph Henry Laboratories of
Physics, Princeton University, Princeton, NJ 08544, USA}
\author{Y. Xia}
\affiliation{Department of Physics, Joseph Henry Laboratories of
Physics, Princeton University, Princeton, NJ 08544, USA}
\author{S.-Y. Xu}
\affiliation{Department of Physics, Joseph Henry Laboratories of
Physics, Princeton University, Princeton, NJ 08544, USA}
\author{D. Qian}
\affiliation{Department of Physics, Joseph Henry Laboratories of Physics, Princeton University, Princeton, NJ 08544, USA}
\affiliation{Department of Physics, Shanghai Jiao Tong University, Shanghai 200030, People's Republic of China}
\author{G. F. Chen}
\author{J. L. Luo}
\author{N. L. Wang}
\affiliation{Beijing National Laboratory for Condensed Matter Physics, Institute of Physics, Chinese Academy of Sciences, Beijing 100080, People's Republic of China}
\author{M.Z. Hasan}
\affiliation{Department of Physics, Joseph Henry Laboratories of
Physics, Princeton University, Princeton, NJ 08544, USA}

\begin{abstract}


We present a polarization and topology resolved study of the low energy band structure in optimally doped superconducting Ba$_{0.6}$K$_{0.4}$Fe$_2$As$_2$ using angle resolved photoemission spectroscopy. Polarization-contrasted measurements allow us to identify and trace all low energy bands expected in models, revealing unexpected symmetry breaking and a surprisingly intertwined Fermi surface topology of hole-like bands near the Brillouin zone center. Band structure correlations across the $\Gamma$-M spin fluctuation wavevector are compared with the superconducting gap anisotropy to suggest a partial scenario for spin-mediated interband instability contributing to superconductivity in the hole doped regime.

\end{abstract}



\date{\today}

\maketitle



Despite intensive efforts, the Fermi surface topology of pnictide high-T$_C$ superconductivity continues to be an unresolved but critically important issue. This is largely due to the fact that it is a multi-band correlated system with nontrivial orbital texture and complicated spin-lattice interactions \cite{discovery,YildirimPhonon,KreyssigCa}. Here we present the low energy dynamics of the optimally doped pnictide superconductor, using comprehensive orbital-polarization resolved band structure mapping to provide an integrated analysis of the orbital symmetries and electron kinetics. Our results reveal a multiplet of Fermi surfaces (FS) of which a pair around the $\Gamma$-point is intricately intertwined with alternating orbital character, and show that there is hole-hole nesting between the pair of outermost Fermi pockets along $\Gamma$-M. This Fermi surface topology and the underlying band dispersions are interpreted in the context of nonlinear electronic correlation-induced corrections to local density approximation (LDA) calculations. Further, the band structure shows unexpected hybridization (symmetry breaking), suggesting the presence of antiferromagnetically ordered domains within the superconducting crystal. By correlating with Fermi surface dependent superconducting gap magnitude measurements, we observe that fine details of the gap structure are supported by ($\pi$,0) interband instability, which supports a spin-fluctuation scenario for the pairing channel \cite{spinOrder}.


\begin{figure}
\includegraphics[width = 7.5cm]{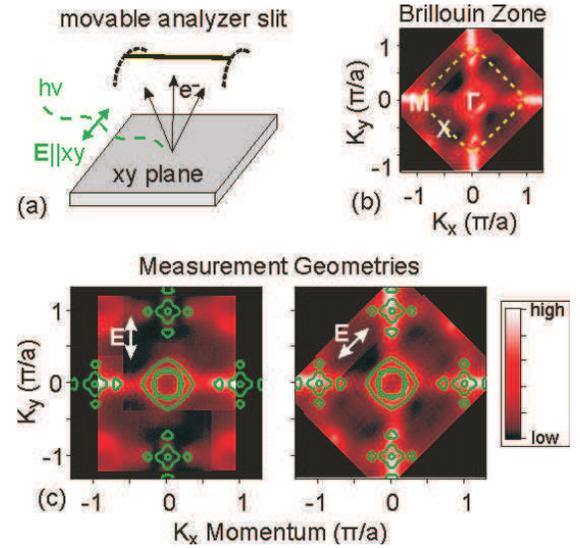}
\caption{{\bf{Complete Fermi Surface polarization mapping}}: (a) Polarization is kept fixed by moving the analyzer rather than the sample during Fermi Surface mapping. (b) High symmetry points in the 2D Brillouin zone are labeled on a Fermi surface map. (c) Fermi surfaces measured at 34 eV incident energy are shown for polarization along the the $\Gamma$-M and $\Gamma$-X high symmetry directions.}
\end{figure}


\begin{SCfigure*}[][t]
\includegraphics[width = 12cm]{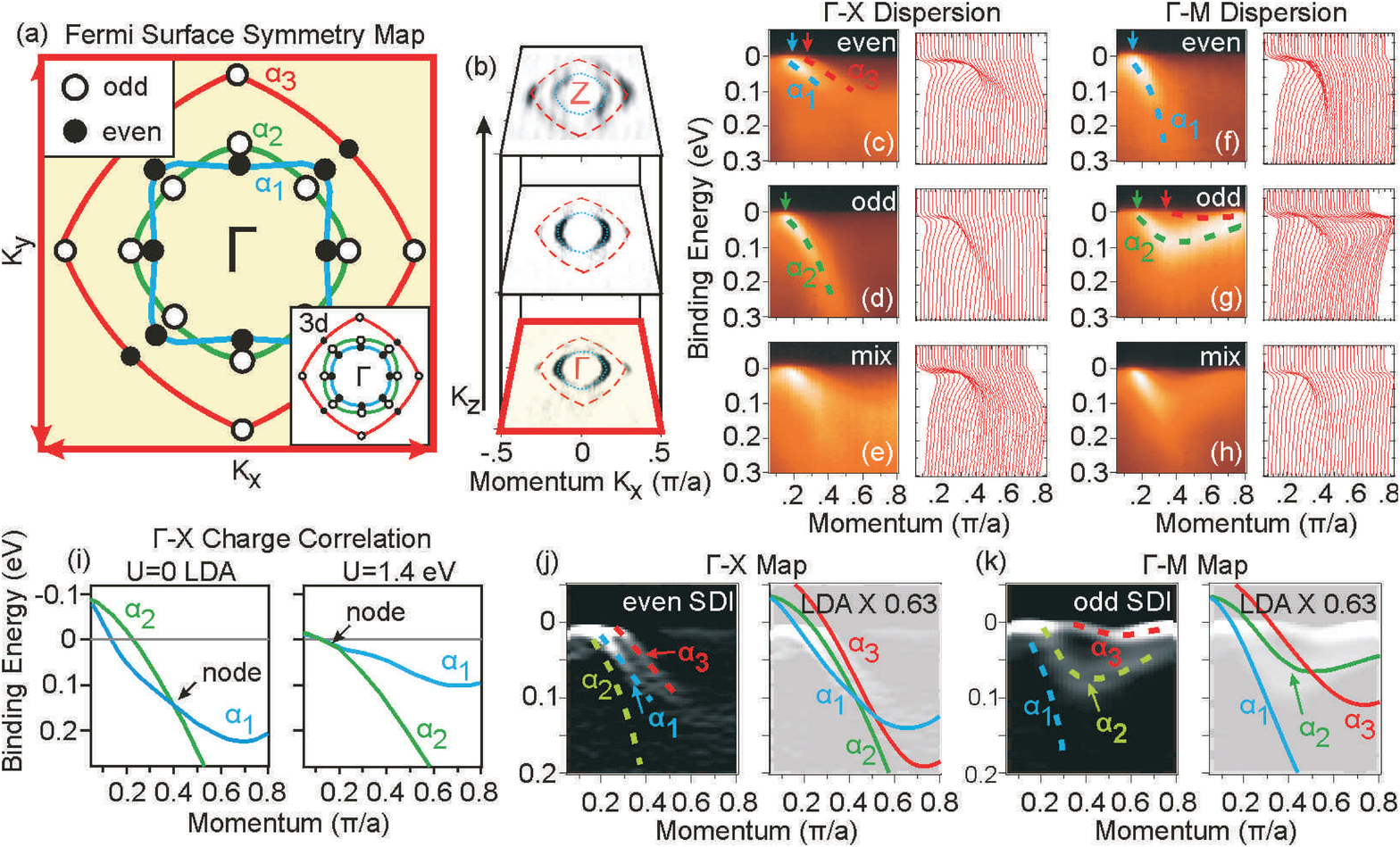}
\caption{{\bf{Band symmetries near the $\Gamma$ point}}: (a) Measured polarization symmetries are labeled on a diagram of the intertwined Fermi surface, and LDA orbital symmetries are summarized in the inset. (b) Momentum dependence of the Fermi surface along the $\hat{z}$ axis, using second derivative images. (c-h) ARPES cuts measured through the 3D Brillouin zone center (K$_z\sim$4 rlu) are shown for the in-plane polarization oriented 0$^o$ (even), 90$^o$ (odd) and at 45$^o$ (mixed reflection symmetry) relative to the cut. (i) Electronic correlation effects raise the G-X Dirac node, beneath which the $\alpha_1$ and $\alpha_2$ Fermi surfaces are intertwined (Numerics from Ref. \cite{AritaRenorm}). (j,k) Second derivative images from panels (c,g) are overlaid with (left) the experimental band structure and (right) LDA dispersions from Ref. \cite{Borisenko} linearly renormalized by a factor of 0.63 and shifted up by 40 meV.}
\end{SCfigure*}

Our previous angle resolved photoemission spectroscopy (ARPES) investigations of iron arsenide band structure have strongly established the importance of incident photon polarization-resolved measurement to view all low energy features \cite{HsiehUD}. In this Letter a more comprehensive approach is adopted to perform separate mappings of band structure and the full Fermi surface with polarization directed along each of the available high-symmetry crystallographic directions. Our presentation of the data begins by focusing near the $\Gamma$-point, using measured reflection symmetries to identify all three low energy bands predicted by LDA. We then trace their dispersion to the M-point to map all low energy bands contributing to the Fermi surface. Other ARPES investigations focusing on the hole-doped iron arsenide Fermi surface (e.g. Ref. \cite{MingYi,Borisenko}) have neglected to perform complete polarization-resolved band mapping, and have not \emph{experimentally traced} the low energy bands along the entire $\Gamma$-M axis, therefor missing significant features.

The optimally doped superconducting Ba$_{1-x}$K$_x$Fe$_2$As$_2$ system was chosen for detailed investigation due to the extremely high sample quality available, as noted in other angle resolved photoemission spectroscopy (ARPES) studies \cite{Wray,DingBS,Borisenko,DingKink,DingGap} and documented in detail in our previous work \cite{Wray} using a combination of scanning tunneling microscopy ($\sim$1$\AA$ rms surface roughness), magnetic susceptibility and ARPES measurements.


\begin{figure*}[t]
\includegraphics[width = 15cm]{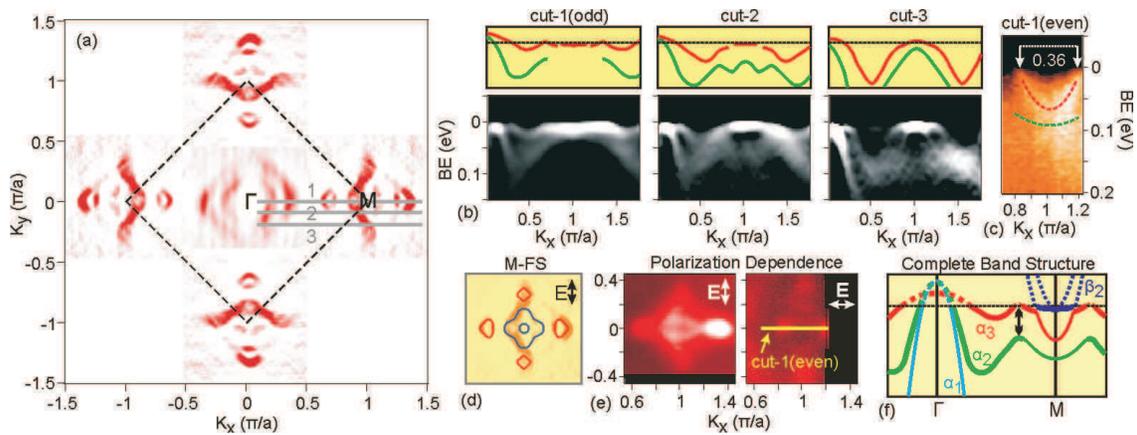}
\caption{{\bf{Band structure near M}}: (a) A symmetrized second derivative image of the Fermi surface is shown near K$_z$=3.5 rlu. (b) Second derivative images of the dispersion cuts labeled on panel (a) are shown with cartoons of the apparent band structure. (c) Two bands near the M-point along the $\Gamma$-M cut are only seen with parallel (even) polarization. (d) A diagram of the M-point Fermi surface is superimposed above data from panel (a). (e) ARPES Fermi surface maps are shown with polarization (left) perpendicular and (right) parallel to the $\hat{x}$ axis. (f) The full experimentally derived $\Gamma$-M dispersion (below E$_F$, K$_z$=3.5 rlu) of all low energy bands predicted by paramagnetic LDA is presented, with thick lines representing the odd symmetry bands visible in cut-1(odd) and thin lines representing the even symmetry sections observed in cut-1(even). The location of a proposed hybridization gap between the $\alpha_2$ and $\alpha_3$ bands is indicated in black.}
\end{figure*}

Momentum space mapping was performed by orienting the high symmetry $\Gamma$-M or $\Gamma$-X crystallographic axis in parallel to incident photon polarization, and moving the analyzer slit to detect ejected electrons with momenta covering a full two dimensional Brillouin zone (Fig-1(a,c)). Out of plane momentum is varied by tuning the incident photon energy. The Brillouin zone mapping convention is labeled in Fig-1(b), with the $\hat{x}$ axis ($\Gamma$-M) defined as the nearest neighbor iron-iron direction. The Fermi surface is composed of three hole-like pockets centered on the $\Gamma$-point and a combination of hole- and electron-like quasiparticle features at the M-point.

Single crystals of Ba$_{0.6}$K$_{0.4}$Fe$_2$As$_2$ (T$_C$=37 K) were grown using the self-flux method \cite{ChenGrowth}. ARPES measurements were performed at the Advanced Light Source beamline 10.0.1 using 34-51 eV photons with better than 10-15 meV energy resolution, respectively, and overall angular resolution better than 1$\%$ of the Brillouin zone (BZ). Samples were cleaved and measured at temperatures below 15 K, in a vacuum maintained below 8$\times$10$^{-11}$ Torr.

Momentum along the $\hat{z}$ axis is parameterized in units of 4$\pi$/c, which has been observed to represent the full periodicity in studies of the electron doped compound BaFe$_{2-x}$Co$_x$As$_2$. An inner potential of 15eV is used to determine K$_z$, consistent with the value used for BaFe$_{2-x}$Co$_x$As$_2$ \cite{Mannella3D}. The linear dimensions of Fermi surface contours, as estimated from Fig-2(b) and similar data near the M-point, increase by roughly 10$\%$ as momentum is varied across the Brillouin zone along the $\hat{z}$ axis, suggesting that the ARPES signal is representative of bulk electronic properties. The full two dimensional Luttinger count ($\frac{FS\,area}{BZ\,area}$) varies from $\sim$18$\%$ hole density in the K$_z$=4 rlu ($\Gamma$-point) plane to $\sim$22$\%$ at the zone boundary (K$_z$=3.5 rlu), in agreement with nominal 20$\%$ hole doping.

Measurements revealing band structure near the three dimensional $\Gamma$-point are shown in Fig-2(c-l). When polarization is 45$^o$ from the cut direction (e.g. $\textbf{E}\parallel\hat{x}+\hat{y}$ for $\textbf{K}\parallel\hat{x}$) the mixed matrix element allows all three hole-like bands surrounding the $\Gamma$-point to be observed simultaneously, however it is difficult to trace them. Cuts parallel or perpendicular to the photon polarization (Fig-2(c-d,f-g)) are observed to selectively suppress at least one band, allowing the remaining bands to be clearly distinguished. Second derivative images in Fig-2(j,k) are used to enhance contrast under geometries for which two bands are visible simultaneously. All bands along the $\Gamma$-X direction follow typical hole-like dispersions, however the two outermost bands in the $\Gamma$-M cut fold upwards as they approach the M-point. The outermost band has a much weaker emission signal than the inner two, but is also less broad, allowing it to show up clearly in the second derivative images.

Measured reflection symmetries at the Fermi surface are summarized in Fig-2(a). Under the scattering geometry used, bands with odd mirror symmetry relative to the momentum cut direction are probed when polarization is perpendicular to the cut, and even symmetry bands are observed when polarization is parallel to the cut. LDA predicts three hole-like bands with dominant 3d$_{xz}$/3d$_{yz}$ (mixed) and 3d$_{xy}$ orbital character near the $\Gamma$-point \cite{KurokiBS}. For a one-to-one identification of these bands with our data, it is necessary to look at details of the reflection symmetry along both high symmetry directions ($\Gamma$-M and $\Gamma$-X).

Along the $\Gamma$-X direction, the reflection symmetry probed by ARPES is also a symmetry of the crystal, meaning that measured bands must be fully symmetric or antisymmetric. The three bands predicted by LDA have even ($\alpha_1$, strong 3d$_{xz}$+3d$_{yz}$), odd ($\alpha_2$, strong 3d$_{xz}$-3d$_{yz}$) and even ($\alpha_3$, strong 3d$_{xy}$) symmetry in this direction. Our data also show two even bands and one odd symmetry band, meaning that the innermost, odd symmetry band can be identified with $\alpha_2$ from LDA.

Along $\Gamma$-M, reflection symmetry is influenced by the corrugated As lattice, allowing some mixing between 3d orbitals that have different reflection symmetries across the Fe-Fe axis. According to LDA calculations, the $\alpha_1$ band has predominant 3d$_{xz}$ character (even), $\alpha_2$ is predominantly composed of 3d$_{yz}$ (odd) and $\alpha_3$ of 3d$_{xy}$ (odd). The band dispersion that most closely matches LDA, with $\alpha_1$ dispersing downwards and $\alpha_2$ and $\alpha_3$ bending upwards (see Fig-2(k)), also matches these d-orbital reflection symmetries. Reconciling band structure along the $\Gamma$-M and $\Gamma$-X axes requires an intertwined Fermi surface, as drawn in Fig-2(a). A more detailed analysis of measurements showing the intersection of $\alpha_1$ and $\alpha_2$ is included as online supplementary information. An intertwined Fermi surface topology can be achieved with selective renormalization (bending) of LDA bands, but not with the sort of global renormalization constant that has been suggested in some recent literature \cite{DingBS,Quazilbash}. New theoretical investigations show that this bending may be caused by charge correlation effects \cite{AritaRenorm}, which raise the energy of the $\alpha_1$-$\alpha_2$ Dirac node (intersection) along the $\Gamma$-X axis, beneath which the band topology is intertwined as we observe (Fig-2(i)).

A second derivative image of the M-point Fermi surface (Fig-3(a,d)) resembles a baseball diamond with outermost dimensions closely matching the size of the outermost $\Gamma$-point Fermi surface ($\alpha_3$ band). In the parallel cuts labeled $\#$1-3, we see that the $\alpha_2$ and $\alpha_3$ bands approach one another closely between the $\Gamma$- and M- points. The point of closest proximity between the bands along arbitrary momentum space cuts is generally well beneath the Fermi level (as seen in cut-2), but approaches the Fermi surface exactly along the $\Gamma$-M axis, at the locations for which hole pockets are observed (cut-1(odd),-3). Rotating polarization to the $\hat{x}$-axis (cut-1(even), even symmetry) reveals the dispersion of two bands near the M-point, one of which is a large electron-like pocket, with a diameter along the $\Gamma$-M axis roughly two thirds as large as the $\hat{x}$ axis separation of the satellite hole pockets (.36$\pi$/a vs. 0.52$\pi$/a), and similar in size to the inner $\Gamma$-point Fermi surface. The location of the other band seen in cut-1(even) is appropriate to connect it with $\alpha_3$.

These observations conflict in several key ways with other experimentally motivated analyses of the band structure. Following the course of the $\alpha_3$ band along $\Gamma$-M reveals that it does fold up to form the outer hole pockets (``propeller" pockets) around the M-point, in agreement with a ($\pi$,0) reconstruction scenario \cite{Borisenko}, and strongly disagreeing with the suggestion that what appear to be propeller pockets are an artifact of intensity from the $\alpha_2$ band \cite{MingYi}. However, our Fermi surface map in Fig-3(a,d) suggests that the inner edge of the propeller pockets is far removed from the electron-like $\beta_2$ band, a situation that does not emerge from ($\pi$,0) reconstruction. The Fermi surface we have traced in Fig-3(d) includes more pockets than are expected in either picture.

One scenario that may self-consistently explain the band maps from cuts $\#$1-3 is shown in Fig-3(f), with heavier lines tracing the bands seen under odd symmetry in cut-1(odd) and thin lines showing the bands only seen under even symmetry (cut-1(even)) or away from the high-symmetry axis (cuts $\#$2-3). Upward bending of the $\alpha_3$ band is suggested to result from hybridization with $\alpha_2$, rather than direct folding across a ($\pi$,0) wavevector from the $\Gamma$-point as in Ref. \cite{Borisenko}. Hybridization between $\alpha_2$ and $\alpha_3$ is not allowed by symmetry in paramagnetic LDA models, but the relevant symmetry can be broken by type-1 antiferromagnetic spin order, which is present in the undoped compound \cite{spinOrder} and may persist in local domains of the superconducting crystal.

Careful tracing of the Fermi surface reveals that the system is too strongly hole doped for there to be perfect ($\pi$,0) nesting between hole- and electron-like Fermi surfaces. However, as shown in Fig-4(a), the outermost ``propeller" hole pockets surrounding the M-point are geometrically nested with the largest hole pocket surrounding the $\Gamma$-point. Because both pockets come from the $\alpha_3$ band, there is no symmetry argument that would prevent \emph{superconducting pairing} between their Fermi surfaces. Nesting between two hole-like Fermi surfaces does not lead to strong spin-fluctuations, and is almost certainly not the primary mechanism for superconductivity. Nonetheless, ($\pi$,0) spin fluctuations present for other reasons, such as exchange induced local spin interactions \cite{AritaLocMom}, could mediate a pairing interaction between the nested hole pockets and allow them to strengthen the superconducting ground state.

\begin{figure}[t]
\includegraphics[width = 8.7cm]{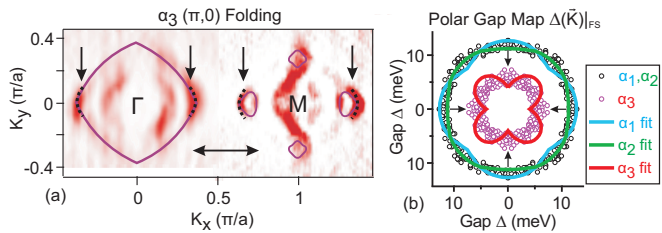}
\caption{{\bf{Interband correlation and gap anisotropy.}} (a) The outermost $\Gamma$- and M-point Fermi surfaces are geometrically nested by a ($\pi$,0) wavevector. (b) Extensive measurements of the superconducting gap $\Delta$(\textbf{K}) from Ref. \cite{DingGap} are compared with fit curves from a cos(K$_x$)$\times$cos(K$_y$) model. Arrows mark points with ($\pi$,0) correlation to satellite hole pockets at the M-point.}
\end{figure}

Such interactions also provide a plausible answer to the long-standing puzzle of why the measured $\alpha_3$ superconducting gap function is isotropic \cite{DingGap,Wray}. In theoretical models of phase shifting s-wave superconductivity, the superconducting order parameter is generally required to disappear midway between the $\Gamma$- and M-points, leading to a reduction in the expected $\alpha_3$ gap size along the $\Gamma$-M axis. A theoretical prediction for the gap distribution based on the lowest order term in such models (cos(K$_x$)$\times$cos(K$_y$)) is drawn with solid lines on a polar plot in Fig-4(b).

The isotropic gap observed in experiments could be strengthened near (but not at) the nodal line by a $\sim$($\pi$,0) pairing interaction with the M-point hole pockets (Fig-4(a)). Unlike hole-electron nesting, interactions between two hole-like Fermi surfaces are robust against hole doping, and could play a role in stabilizing superconductivity through the large part of the phase diagram in which overdoped superconductivity is found. ($T_C>0$ for $0.1<x\leq1$ at 1 atm pressure \cite{ChenDoping,RotterDoping}) Based on their low Fermi velocities and small size at x=0.4 doping, it is possible that the emergence and ($\pi$,0) interband instability of the M-point satellite hole pockets are characteristics that distinguish optimally doped superconducting Ba$_{1-x}$K$_x$Fe$_2$As$_2$ from underdoped crystals.



The results presented in this paper strongly emphasize the need for more sophisticated first principles numerical modeling that can comprehensively address short range spin order and ARPES matrix elements. Our data show several features that are not present in typical paramagnetic LDA calculations, such as the intertwined $\Gamma$-point Fermi surface and hole pockets near the M-point apparently resulting from hybridization of the $\alpha_2$ and $\alpha_3$ bands. We note that interactions between spin and the lattice dimensions are unusually strong for iron pnictides, and symmetry breaking spin correlations in the pnictide plane are necessary to reconcile first principles theories with the crystal structure and phonon spectrum \cite{YildirimPhonon}. With respect to experimental methodology, we emphasize that due to the distribution of matrix elements and the weak photoemission intensity of some bands, a detailed, comprehensive comparison with LDA is only possible if the band dispersions are separately traced. It is not sufficient to simply establish a correlation between the LDA dispersions and regions of intensity in the ARPES spectrum.

In summary, we present a polarization resolved ARPES study of the Fermi surface and band structure in optimally doped Ba$_{1-x}$K$_x$Fe$_2$As$_2$. We observe the dispersion of three hole-like Fermi sheets surrounding the $\Gamma$-Z axis and demonstrate that two of them intersect, providing the finely resolved Fermi surface topology in that region of momentum space and emphasizing the importance of electronic correlation effects in shaping band structure near the Fermi level. Polarization-symmetry characterized mapping enriches this picture with respect to theoretical models, by revealing the momentum space distribution of band symmetries. Finally, by tracing band structure along the $\Gamma$-M axis, we identify geometrical nesting and apparent hybridization related to $\sim$($\pi$,0) spin fluctuations, and compare these observations with the superconducting gap size to support a scenario of interband interactions.


\begin{acknowledgments}

We acknowledge conversations with Ying Ran, Ashvin Vishwanath, B.A. Bernevig, Takami Tohyama, Z. Tesanovic, Igor Mazin and Fa Wang.

\end{acknowledgments}

\end{document}